\documentclass[aps,prl,twocolumn,groupedaddress,superscriptaddress,showpacs]{revtex4}
\usepackage{graphicx}

\begin{document}

\title{Edge effects in Bilayer Graphene Nanoribbons}

\author{Matheus P. Lima}
\email[]{mplima@if.usp.br}
\affiliation{Instituto de F\'isica,
Universidade de S\~ao Paulo, CP 66318, 05315-970, S\~ao Paulo, SP,
Brazil.}

\author{A. Fazzio}
\email[]{fazzio@if.usp.br}
\affiliation{Instituto de F\'isica,
Universidade de S\~ao Paulo, CP 66318, 05315-970, S\~ao Paulo, SP,
Brazil.} \affiliation{Centro de Ci\^encias Naturais e Humanas,
Universidade Federal do ABC, Santo Andr\'e, SP, Brazil}

\author{Ant\^onio J. R. da Silva}
\email[]{ajrsilva@if.usp.br}
\affiliation{Instituto de F\'isica,
Universidade de S\~ao Paulo, CP 66318, 05315-970, S\~ao Paulo, SP,
Brazil.}

\date{\today}

\begin{abstract}
We show that the ground state of zigzag bilayer graphene
nanoribbons is non-magnetic. It also possesses a finite gap, which
has a non-monotonic dependence with the width as a consequence of
the competition between bulk and strongly attractive edge
interactions. All results were obtained using {ab initio} total
energy density functional theory calculations with the inclusion
of parametrized van der Waals interactions.
\end{abstract}

\pacs{73.22.-f, 72.80.Rj, 61.48.De, 71.15.Nc}

\maketitle

Since the synthesis of graphene\cite{gr1}, a plethora of
intriguing properties has been found in this two dimensional zero
gap crystal due to the presence of massless fermions with a high
mobility\cite{gr2,gr3}. Besides the monolayer, stacking two layers
of graphene still preserves the high mobility, and some features
of the electronic spectrum can be controlled, for example, by
applying an external electric field\cite{bil}. Measurements of
quantum hall effect and quasi-particle band structure indicate
qualitative differences between monolayers and bi-layers. The
occurrence of this rich physics at
room-temperature\cite{room1,room3} has attracted a great interest
in designing graphene-based nanoelectronic devices. In this
scenario, it is fundamental to establish and control an energy gap
($E_g$).

A possibility is to introduce lateral quantum confinement via
synthesis of single layer (GNR)\cite{nrgaps1,nrgaps2} or bilayer
(B-GNR)\cite{lam2008,castro2008} graphene nanoribbons by plasma
etching or chemical routes\cite{nr1,nr2}. This opens a gap that
rises the possibilities to use graphene in nanoelectronics, where
small widths (sub-10 $nm$) is required for room temperature
applications\cite{room3}. However, the BGRNs are less sensitive to
external perturbations in comparison with GNRs, and hence, they
may be more appropriate to fabricate hight quality
nano-devices\cite{noise}.

The electronic structure of the nanoribbons, including the gap,
are largely affected by the geometric pattern (zigzag or armchair)
at their edges. GNRs with zigzag edges, in particular, have as a
distinct feature the presence of edge states that introduce a
large density of states (DOS) at the Fermi energy. Theoretical
works predict that this configuration is unstable, and there will
be the appearance of a magnetic order that leads to a removal of
this large DOS peak\cite{nrgaps1,nrgaps2,nrgaps3}. Magnetism in
GNRs has been intensively investigated as a possible way to
develop spintronic devices\cite{spintr1,spintr2,spintr3}. An
anti-ferromagnetic (ferromagnetic) order between the two edges
leads to a semiconductor (metallic)
state\cite{nrgaps1,nrgaps2,nrgaps3}. It is also believed that
magnetism is necessary to open a gap in zigzag B-GNR
(B-ZGNR)\cite{macdonald}.

In this Letter we investigate the geometrical and electronic
structure of B-ZGNR, and show that the ground state of these
systems is {\it non-magnetic and possesses a non-monotonic finite
gap}. There are two possible edge alignments for the B-ZGNR,
called $\alpha$ and $\beta$ (see Fig. 1). We found that the
$\alpha$ alignment is energetically favorable, with an inter-layer
edges attraction, whereas for the $\beta$ alignment there is an
inter-layer edges repulsion. These edge-related forces cause a
deviation from the exact Bernal stacking, resulting in a
non-monotonic behavior of the energy gap with the width $w$ for
the $\alpha$ B-ZGNR, with a maximum value at $w\approx 3.5nm$.
These results differ qualitatively from GNRs with zigzag edges
(M-ZGNR)\cite{nr1,nr2}

\begin{figure}[b]
\includegraphics[width=8.5cm]{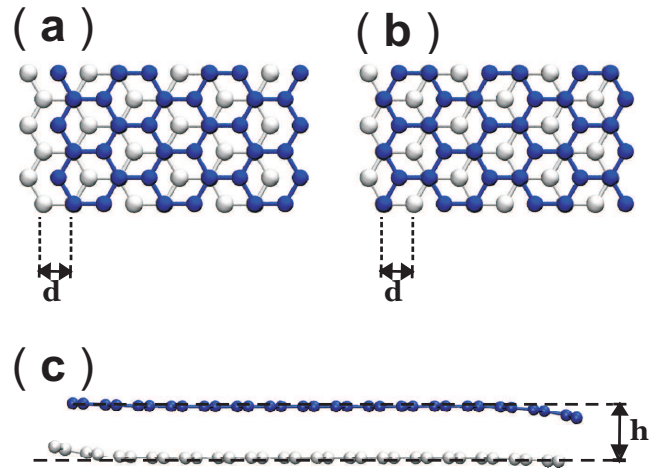}
\caption{\label{fig1} Bilayer graphene nanoribbons with (a)
$\alpha$ and (b) $\beta$ edge alignment. (c) Side view of bilayer
graphene nanoribbons. The blue (white) atoms form the upper
(bottom) layer.}
\end{figure}

All our results are based on {\it ab initio} total energy Density
Functional Theory\cite{dft} (DFT) calculations. In order to
correctly describe multi-layer graphitic compounds, it is
necessary to include van der Waals (vdW) interactions. The use of
fully relaxed total energy DFT calculations to study such systems
suffers from serious limitations, since the most traditional
exchange correlation ($xc$) functionals in use today do not
correctly describe these terms. With the LDA $xc$, the geometry is
correctly described but the inter-layer binding energy is
underestimated by 50\%, whereas the GGAs $xc$ do not even
correctly describe the geometrical
features\cite{graphite_cohesion}. Thus, in order to be able to
investigate the geometries and relative energies of B-ZGNRs, we
include a non-local potential in the Kohn-Sham (KS) equations that
correctly describes the vdW interactions. We modified the SIESTA
code\cite{siesta} adding in the KS Hamiltonian\cite{dcap-nosso}
the dispersion corrected atom centered potential
(DCACP)\cite{dcacp1}. This correction is sufficiently accurate to
describe weakly bonded systems\cite{dcacp3} with the vdW
interactions included in the whole self-consist cycle, providing
accurate values for both forces (and thus geometries) and total
energies. Our implementation was successfully tested (see Table
\ref{tab1}), and was employed to obtain the results here
reported\cite{parm,pbe,tm}.

\begin{table}
\caption{\label{tab1} Inter-layer binding energy ($E_b$), in
$eV/atom$, and distance $h$, in {\AA}, for graphite and a graphene
bilayer, which is not bound (NB) at the GGA level.}
\begin{ruledtabular}
\begin{tabular}{lccccc}
         &    & present  &  Exp  &   LDA  & GGA(PBE) \\ \hline
Graphite &\begin{tabular}{c} $E_b$ \\ h \end{tabular}  &
          \begin{tabular}{c} 0.054 \\ 3.350 \end{tabular}  &
          \begin{tabular}{c} $0.052\pm 0.005$\tablenote{From Ref. \cite{prb69.155406}} \\ $3.356$\tablenote{From Ref. \cite{prb75.153408}}\end{tabular}  &
          \begin{tabular}{c} 0.030 \\ 3.200 \end{tabular}   &
          \begin{tabular}{c} 0.003 \\ 4.5 \end{tabular}     \\
     &  &  &  &  \\
Bilayer  &\begin{tabular}{c} $E_b$ \\ h \end{tabular}  &
          \begin{tabular}{c} 0.027 \\ 3.320\end{tabular} &
          \begin{tabular}{c} - \\ - \end{tabular}  &
          \begin{tabular}{c}  0.017 \\ 3.202 \end{tabular} &
          \begin{tabular}{c}  NB \\ NB \end{tabular}        \\
\end{tabular}
\end{ruledtabular}
\end{table}

We investigated B-ZGNR composed by two M-ZGNR passivated with
hydrogen, and widths\cite{width} that range from $w=0.6$ to
$w=4.5$ $nm$. The layers are in the Bernal stacking, which means
that there are two types of C atoms, those that are positioned
above the center of the hexagons of the other layer, defining a
B-sublattice, and those right on top of the C atoms of the other
layer, forming an A-sublattice. An infinite graphene bilayer has
no gap, and the orbitals at the Fermi level are located at the
B-sublattice. When we cut the layer along the zigzag edge, there
are two possible alignments (Fig.\ref{fig1}): (a) the $\alpha$
alignment, where the outermost edge atoms belong to the
A-sublattice, and (b) the $\beta$ alignment, where the outermost
edge atoms belong to the B-sublattice. Thus, only the inter-layer
edge interaction differs. Two geometrical distortions have proven
to be important: (i) an edge distortion that causes a curvature in
the ribbons (see Fig.\ref{fig1}(c)), and (ii) a lateral deviation
from the perfect Bernal stacking. To quantify this deviation, we
define the quantity $u\equiv d_{C-C}-d$, where $d$ is shown in
Fig.\ref{fig1}(a) and (b), and $d_{C-C}$ is the carbon-carbon bond
length. The perfect Bernal stacking corresponds to $u=0$.

\begin{figure}[h]
\includegraphics[width=8.5cm]{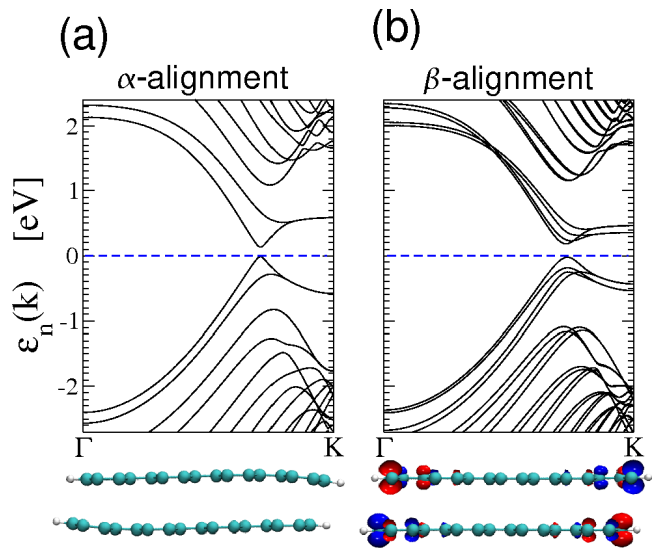}%
\caption{\label{fig2} Ground State of fully relaxed B-ZGNRs
generated by stacking two (5,0) M-ZGNR. Below each band structure
the geometry and local magnetization are presented. (a) $\alpha$
alignment. This state is non-magnetic and presents a geometric
distortion near the edge. (b) $\beta$ alignment. This state shows
an AF in-layer and AF inter-layer magnetic order.}
\end{figure}

The geometries and band structures of fully relaxed B-ZGNRs with
$w\approx 1.0~nm$ for $\alpha$ and $\beta$ alignments are presented
in Fig. \ref{fig2}. In B-ZGNR with the $\beta$ alignment,
similarly to M-ZGNRs, a non spin-polarized calculations leads to a
high DOS at the Fermi energy, and a magnetic order is required to
split these localized edge states at the $K$ symmetry point. In
order to establish the possible spin polarized configurations, we
used four initial guesses for the density matrix before starting
the self-consistency cycle, which are: $i$) anti-ferromagnetic
(AF) in-layer and inter-layer, $ii$) ferromagnetic (F) in-layer
and inter-layer, $iii$) AF in-layer and F inter-layer and $iv$) F
in-layer and AF inter-layer. As well as non-polarized
calculations. From all calculations, the AF in-layer and
inter-layer guess leads to the lower energy state (Fig.
\ref{fig2}(b)). However, the energy differences are less than
$k_BT$\cite{macdonald}.

At the $\alpha$ alignment (Fig. \ref{fig2}(a)), on the other hand,
we obtain a qualitatively new situation. There is a strong
attractive interaction between the edge atoms of the two layers,
with a resulting geometric distortion that decreases the distance
between them (Fig. \ref{fig2}(a)). For all $\alpha$ B-ZGNRs that
we have investigated, the final geometry always had an inter-layer
edge atoms distance around $3.0$ $\mathring{A}$. The final
configuration is non-magnetic and with a finite gap, contrary to
previous results where the presence of a gap was intrinsically
coupled to a magnetic state\cite{macdonald}. Note that if we do
not allow the atoms at the two layers to relax, but simply
optimize the inter-layer distance ({\it i.e.}, the layers keep
their planar geometries), a magnetic configuration is still
necessary to open a gap\cite{macdonald}. However, this
configuration has higher energy.

If we take one of the mono-layers that form the final relaxed
$\alpha$ B-ZGNR, and perform a calculation without letting the
atoms relax, we obtain an energy increase, when compared to the
lowest energy M-ZGNR (AF in-layer\cite{spintr1}), that can be
broken down in two components. Considering M-ZGNR with widths
larger than 2 $nm$, if we allow the distorted monolayer to be
magnetic, the energy increase is $\approx 0.1~eV/nm$, which can be
viewed as the elastic contribution. If we now consider a
non-magnetic configuration for this distorted monolayer, which is
the situation in the $\alpha$ B-ZGNR, there is an extra energy
increase of $\approx 0.4~eV/nm$, {\it i.e.}, an overall energy
penalty of $\approx 0.5~eV/nm$. Considering the two monolayers,
the total energy cost to deform and demagnetize the $\alpha$
B-ZGNR is $\approx 0.96~eV/nm$. This energy increase is more than
compensated by the edge atoms interaction, and the energy gain is
in part associated with a large split of the localized edge
states. At the $\beta$ B-ZGNR (Fig. \ref{fig2}(b)), the unique way
to diminish the DOS associated with the localized edge states at
the Fermi energy is via a magnetic ordering, and the system tends
to increase the inter-layer edge atoms distance in order to allow
a bigger magnetization, given rise to a repulsive inter-layer edge
interaction.

\begin{figure}[h]
\includegraphics[width=8.5cm]{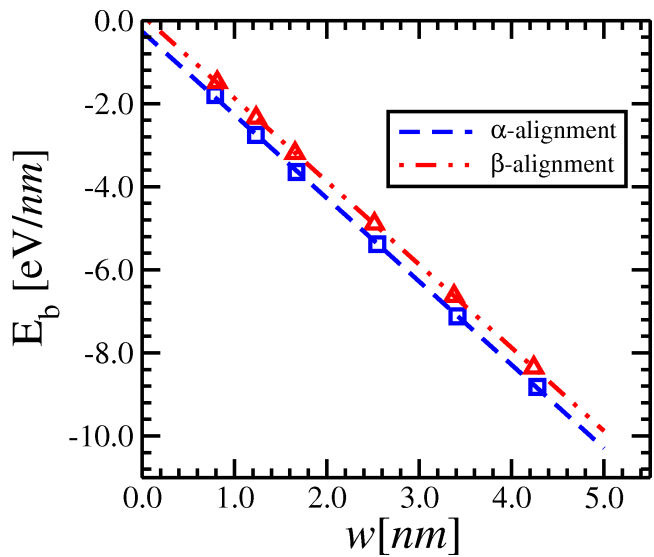}%
\caption{\label{fig3} Dependence of the binding energies with the
width.}
\end{figure}

Comparing the two alignments, the $\alpha$ B-ZGNR results
energetically favorable. This is an even more important conclusion
considering that most of the calculations used the $\beta$
B-ZGNR\cite{castro2008,bifita-tb}. Fig. \ref{fig3} presents the
dependence of the B-ZGNRs binding energies, for both edge
alignments, as a function of $w$ (calculated relative to two
isolated lowest energy M-ZGNR). The interaction between the layers
can be separated into two components: $i$) edge interactions, that
do not depends on the width, $ii$) and bulk interactions, that
increase linearly with $w$. The binding energies (per unit length)
can be well adjusted with:
\begin{eqnarray}
E_b(w)=a+bw.
\end{eqnarray}
Since there are two edges, $a/2$ is the inter-layer edge
interaction energy per unit length, and $b$ is the inter-layer
bulk interaction energy per unit area. At the $\alpha$ alignment,
$a=-0.26eV/nm$ and $b=-2.0eV/nm^2$, indicating that the edges
interaction is attractive ($a<0$). For the $\beta$ alignment,
$a=+0.13eV/nm$ and $b=-2.0eV/nm^2$, indicating that there is a
repulsive edges interaction ($a>0$), showing that its stability
results solely from the bulk. The parameter $b$, as expected, does
not depend on the edge alignment, and it is very close to our
calculated bulk inter-layer interaction in a graphene bilayer
(0.027 $eV/atom$=$-1.99 eV/nm^2$).

\begin{figure}
\includegraphics[width=8.5cm]{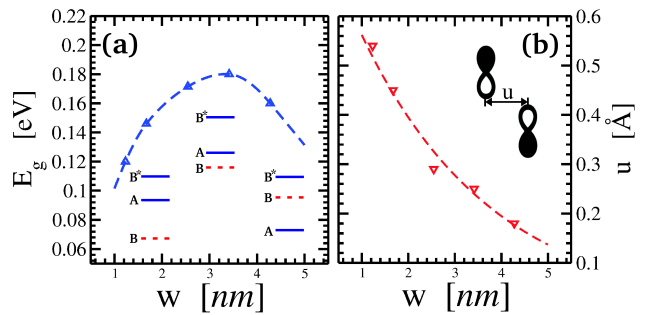}%
\caption{\label{fig4} Dependence of the (a) energy gap (inset
illustrates the change of character of the VBM), and (b) the
lateral deviation $u$ with the width $w$ (inset indicates how $u$
affects the inter-layer A-sublattice $p$-orbitals interaction).}
\end{figure}

For the $\alpha$ B-ZGNR, there is a competition between the forces
deriving from the bulk, that do prefer the Bernal pattern of
stacking, and the forces deriving from the edges, that tend to
maintain the inter-layer edge carbon atoms distance close to
3~{\AA}. The system then minimizes the overall energy penalty by
simultaneously optimizing both the deviation $u$ from the exact
Bernal stacking, and the elastic energy associated with the
ribbon's curvature. Thus, for narrower B-ZGNR the system prefers
to have a larger value of $u$ and a smaller overall curvature
whereas for wider B-ZGNR the deviation $u$ tends to decrease to
minimize the bulk energy penalty, since now it is possible to have
a softer curvature that is somewhat localized at the edges when
compared to the total width of the ribbon (see Fig.
\ref{fig1}(c)). As a result, the dependence of the lateral
deviation $u$ with the width $w$ is well adjusted by
$u(w)=0.80e^{-0.35w}$ (with $u$ in [$\AA$], and $w$ in [$nm$]).
Moreover, as a result, the carbon-carbon bond lengths do not
significantly differ from their values in the M-ZGNRs. For this
situation, the average inter-layer distance $h$ is close to its
value in the graphene bilayer (see Tab. \ref{tab1}).

Without the geometrical deformation caused by the inter-layer edge
interactions, a monotonic decrease of the energy gap is expected
due to the quantum confinement ($\propto
1/w$)\cite{nrgaps2,macdonald}. However, for small ribbons, we find
that the character of the valence band maximum (VBM) is located at
the A-sublattice, as opposed to larger ribbons where it is located
in the B-sublattice (see Fig. \ref{fig4}(a)), similarly to the
Fermi level orbitals in the infinite graphene bilayer. Moreover,
since the interaction between the C atoms in the A-sublattice
increases when $u$ decreases (see Fig.\ref{fig4}(b)), the gap
initially increases with $w$. However, for $w\gtrsim3.5~nm$, due
to the quantum-confinement decrease, there is a crossover between
the two highest occupied bands, and the character of the VBM is at
the B-sublattice. Thus, this leads to a non-monotonic behavior of
the energy gap with $w$ (Fig. \ref{fig4}(a)).

For the $\beta$ B-ZGNR, there is a repulsive interaction between
the edges, in such way that the inter-layer edge carbon atoms
distance is close to 3.7~{\AA}. There happens a small negative
lateral deviation ($u<0$) that can be neglected when $w>1.6nm$.
Note that for the $\alpha$ B-ZGNR, due to the attractive edges
interaction, $u>0$. We also found that, despite the presence of a
magnetic order, the energy gap disappears when $w>3~nm$.

Summarizing, we unequivocally show that for B-ZGNR the edge
alignment $\alpha$ is the lowest energy configuration. This is a
result of the strong attractive interaction between the edges,
that is manifested in an observed chemical bonding between the
inter-layer edge carbon atoms, and that significantly influences
the geometry and electronic structure of bilayer nanoribbons with
{\it sub-10} $nm$ widths. As a consequence, the ground-state is
non-magnetic and possesses a finite gap, which presents a
non-monotonic dependence with the width.

We acknowledge helpful discussions with M. D. Coutinho-Neto
regarding the DCACP, financial support from FAPESP and CNPq.


\begin{thebibliography}{99}
\bibitem{parm} We use a generalized gradient
aproximation\cite{pbe}, norm-conserving pseudopotentials\cite{tm},
a mesh cut off of $400Ry$ for the grid integration, a $1\times
1\times 50$ mesh for the first Brillouin zone sampling, a force
criterion of $10meV/\mathring{A}$, and a supercell approximation
with a lateral separation between the images of $20 \mathring{A}$.
\bibitem{width}
We have defined the width as the lateral distance between the
outermost C edge atoms of the bilayer.
\end{thebibliography}

\end{document}